\documentclass[a4paper]{article}
\usepackage{amsfonts}
\usepackage{amssymb}
\usepackage[margin=2cm]{geometry}
\usepackage[dvips]{graphicx}

\usepackage[T1]{fontenc}
\usepackage[latin2]{inputenc}

\usepackage{amsmath}
\usepackage{bbm}

\begin{document}

%\draft
%\twocolumn

\title{False qubits II. Entanglement of Josephson junctions}

\author{Robert Alicki \\ 
  {\small
Institute of Theoretical Physics and Astrophysics, University
of Gda\'nsk,  Wita Stwosza 57, PL 80-952 Gda\'nsk, Poland}\\
}

\date{\today}
% \date{July 26, 2003}
\maketitle

\begin{abstract}
The recent experimental evidence for entangled states of two Josephson junction qubits \cite{MS} is briefly discussed.
It is argued that the interpretation of the experimental data strongly depends on the assumed theoretical model. Namely, the qubit states
are supposed to be the lowest lying eigenstates of a certain effective Hamiltonian and hence automatically orthogonal, while the simple analysis within a more fundamental many-particle model shows that those states should strongly overlap.
This makes the standard interpretation of the measurement procedure questionable.

\end{abstract}

In the recent paper of the author \cite{RA} it has been argued that the Josephson junction (JJ) is essentially a classical system and therefore cannot serve as an implementation of a qubit. On the other hand the very recent experiment on two coupled phase  JJ strongly supports the idea of superconducting qubits which can be prepared in relatively stable entangled states \cite{MS}. To resolve this paradox we compare two models of the JJ (for simplicity the Cooper pair box). The first one is a phenomenological model usually based on the formal quantization of the Kirchhoff's equation for the suitable superconducting circuit \cite{D},\cite{W} but here derived from a certain simplified quantum Hamiltonian via semiclassical limit
and "requantization" of the relevant degree of freedom \cite{RA}. The obtained system is a nonlinear quantum oscillator for which
its two lowest eigenstates define a qubit.  The (destructive) qubit measurement is done by applying a strong pulse such that the higher energy state (say $|1\rangle$) tunnels out of the potential well. If it happens then it is concluded that the qubit was in the state $|1\rangle$ otherwise in the state $|0\rangle$. This interpretation is based solely on the assumption that the only possible outcomes are two orthogonal eigenstates $|0\rangle$,$|1\rangle$.
The second model  based on the approximated form of the  Bose-Einstein condensate's wave functions which should correspond to $|0\rangle$,$|1\rangle$  exhibits a strong overlap $\langle 0|1\rangle \simeq 1$.
\par
\emph{The Cooper pair box}
We begin with the simplified model of a JJ assuming that the Cooper pairs can be treated as a free bosonic gas
below the critical temperature of Bose-Einstein condensation. We have two electrodes ("1" and "2") made of a superconducting material separated by a thin layer of an insulator which allows for tunneling of Cooper pairs.
The annihilation and creation operators  $a_1, a_1^{\dagger}$ and $a_2 , a_2^{\dagger}$  correspond to the
ground states of a boson (Cooper pair) in separated electrodes.
In order to construct a superconducting qubit one tries to supress the tunelling of many Cooper pairs by using the circuit consisting of a small superconducting island "1" connected  via a JJ to a large superconducting reservoir "2". Coulomb
interaction between Cooper pairs in a small electrode become important and   can be modelled by  the quadratic term  in the Hamiltonian below, with ${\bar n}_1$ being the background number of Cooper pairs in the island corresponding to a neutral reference state. The second term contains potential difference $U$ between electrodes and the term proportional to $\lambda$ describes the tunneling
of Cooper pairs
\begin{equation}
H = E_C( a_1^{\dagger}a_1 - {\bar n}_1 )^2 +\frac{U}{2}( a_1^{\dagger} a_1 - a_2^{\dagger}a_2)  + \frac{\lambda}{2} (a_1 a_2^{\dagger} + a_1^{\dagger}a_2)\ .
\label{nJJ}
\end{equation}
\par
The standard picture of the Cooper pair box can be obtained from the Hamiltonian (\ref{nJJ}) treating the operators
$a_1, a_2$ as classical variables and using the following formal
substitutions 
\begin{equation}
n_1 = a_1^{\dagger}a_1,\ n_2=a_2^{\dagger}a_2,\  n= n_1-{\bar n}_1 ,\ a_1= \sqrt{n_1}e^{-i\phi_1},\ a_2= \sqrt{n_2}e^{-i\phi_2} ,\ \phi_1 -\phi_2=\phi
\label{cl}
\end{equation}
which (under the condition
$n << n_1, n_2, N\equiv n_1 +n_2$) yields 
the Hamiltonian (up to an irrelevant constant)  \cite{D},\cite{W}
\begin{equation}
H = E_C(n- n_g )^2 - E_J \cos\phi\ \ ,\ n_g=  -\frac{U}{2E_C}\ ,\ E_J= -\frac{\lambda}{2}\sqrt{(N-{\bar n}_1){\bar n}_1}\ .
\label{JJham1}
\end{equation}
The next step in the standard approach is to "requantize" the variables $n, \phi$ assuming the following
canonical commutation relations \cite{D},\cite{W}
\begin{equation}
[\phi , \pi]=i \ ,\ \pi = n - n_g
\label{ccr}
\end{equation}
and to consider the properties of the "quantum nonlinear oscillator" with the Hamiltonian (\ref{JJham1}). The two lowest lying eigenstates, which are automatically orthogonal, are supposed to provide the superconducting implementation of a qubit.
\par
\emph{Many-particle wave functions}

The fundamental property of the Bose-Einstein condensate is the form of its N-boson wave function which is approximatively given by a tensor product of the identical single-boson wave functions $\psi$ \cite{KH}. Therefore, the set of condensate's wave functions does not form a linear subspace in the Hilbert space of the N-particle system.
Observed "coherent superpositions" of two Bose condensates are described by the states of the form 
\begin{equation}
\Psi = \otimes_{N} (c_1\psi_1 + c_2\psi_2) 
\label{wfcond}
\end{equation}
which are not superpositions of states corresponding to separated parts of the condensate. Moreover, even on the single-boson level the "superposition principle" leading to $c_1\psi_1 + c_2\psi_2$ holds only approximatively as the single-boson function satisfies a \emph{nonlinear} Gross-Pitaevski equation of motion.
\par
Due to the coherent tunneling of Cooper pairs the wave function $\psi= c_1\psi_1 + c_2\psi_2$ is a superposition of two wave functions localized in the first and the second electrode, respectively. In particular,  $N$ Cooper pairs distributed among two electrodes
with $N_1$ pairs in the first and $N_2= N-N_1$ in the second one are described by the state
\begin{equation}
\Psi = \bigotimes_{N} \Bigl(\sqrt{\frac{N_1}{N}}\varphi_1 + \sqrt{\frac{N_2}{N}}\varphi_2\Bigr) 
\label{wf}
\end{equation}
where $\varphi_1$ and $\varphi_2$ are  orthogonal and normalized single-boson wave functions corresponding to ground states in electrodes
"1" and "2", respectively. The states $|0\rangle$,$|1\rangle$ of the Cooper pair box qubit in the standard picture differ by a small number $\delta N << N_1$ of  Copper pairs in the electrode. Then in the many-body picture they can be written as
\begin{equation}
|0\rangle\equiv \Psi_-\ ,\ |1\rangle\equiv \Psi_+\ ,\Psi_{\pm} = \bigotimes_{N} \Bigl(\sqrt{\frac{N_1\pm \delta N/2}{N}}\varphi_1 + \sqrt{\frac{N_2\mp \delta N/2}{N}}\varphi_2\Bigr) 
\label{wf1}
\end{equation}
Their scalar product, under the assumption $N_1 << N_2$, reads
\begin{equation}
\langle\Psi_- |\Psi_+\rangle  \simeq \bigl( 1 - \frac{(\delta N)^2}{8N_1 N}\bigr)^N \simeq e^{-(\delta N)^2/8N_1}\simeq 1-(\delta N)^2/8N_1\ .
\label{sp}
\end{equation}
It follows that all states of the condensate which differ by a "microscopic number" of Cooper pairs in the electrode
form a narrow cone (when depicted as "rays") and we cannot find among them two orthogonal ones.
\par
\emph{Conclusion}
We have compared two descriptions of the JJ qubit states: first based on the standard phenomenological effective Hamiltonian and the second applying many-particle wave functions of the Bose-Einstein condensate. The results are completely different and because the second approach seems to be more fundamental the question of the validity of the standard interpretation of the measurement data in JJ qubit experiments is justified.

\end{document}